\documentclass[11pt,twoside]{article}
\usepackage{./asp2014}

\aspSuppressVolSlug
\resetcounters

\newcommand{\fsub}{\ifmmode f_{\rm sub}\else$f_{\rm sub}$\fi}
\newcommand{\msub}{\ifmmode M_{\rm sub}\else$M_{\rm sub}$\fi}
\newcommand{\mlim}{\ifmmode M_{\rm lim}\else $M_{\rm lim}$\fi}
\newcommand{\msun}{\ifmmode M_\odot\else $M_\odot$\fi}

\begin{document}

\title{Probing dark matter and galaxy evolution with proper motions of galaxies}
\author{Thomas J. Maccarone,$^1$  Anthony H. Gonzalez$^2$}
\affil{$^1$ Texas Tech University, Lubbock TX USA; \email{thomas.maccarone@ttu.edu}}
\affil{$^2$ University of Florida, Gainesville, FL, USA}


\paperauthor{Sample~Author2}{Author2Email@email.edu}{ORCID_Or_Blank}{Author2 Institution}{Author2 Department}{City}{State/Province}{Postal Code}{Country}
\paperauthor{Sample~Author3}{Author3Email@email.edu}{ORCID_Or_Blank}{Author3 Institution}{Author3 Department}{City}{State/Province}{Postal Code}{Country}

\begin{abstract}

We discuss several problems that can be solved by measuring the proper
motions of galaxies in clusters of galaxies, something that should be
feasible in the Next Generation Very Large Array era.  First, in the
Virgo Cluster, the Laser Interferometric Space-based Antenna should
provide standard siren distances for a substantial number of galaxies
which contain black hole-black hole binaries.  Measuring the proper
motions of these galaxies will give 6-dimensional phase space
information about the orbits of these galaxies, providing a unique set
of tracers of cluster dynamics.  Additionally, in other clusters
proper motion measurements should allow the determination of whether
spiral galaxies are on their first infall into the cluster, and
measurements of the proper motions of galaxies in merging clusters
will allow direct measures of the speed of the merger, which can then
be compared with estimates from shocks.

\end{abstract}

\section{Description of the problem}

Understanding the nature of the dark matter and how it impacts
structure formation is a key problem in astrophysics.  On the scales
of individual galaxies, rotation curves can provide good
understandings of the relationship between distance from the center of
mass of the galaxies, and the mass enclosed, albeit with some
uncertainty about deviations from sphericity of the Galactic halo.
Over the next $\sim15$ years we can expect to see the launch of the
LISA mission, which will have the sensitivity to detect stellar-mass
double black holes in the Virgo Cluster of galaxies, and from which we
expect to see a non-negligible number of sources
\citep{Kremer18,Jin18}.  These sources should have positional
accuracies on the order 1 square degree and, crucially, standard siren
distances accurate to about 1\%, allowing them to be associated
uniquely with their host galaxies.  In other contexts, the proper
motions of galaxies alone, even without the precise distances expected
from gravitational wave sources, should allow fundamental tests of the
dark matter hypothesis in nearby groups.  Even without standard siren
distance measurements, proper motions should yield useful estimates of
the merger speed in merging clusters of galaxies, and determination of
whether galaxies in clusters are on their first infalls, and how this
might correlate with galaxy morphology and gas content.


\section{Scientific importance and astronomical impact}

We are now at the start of an era in which Gaia is revolutionizing our
understanding of the structure of the Milky Way galaxy by obtaining
6-D phase space measurements for a large number of objects.  At the
present time, no such information is available for any galaxies in
clusters of galaxies, or even in nearby groups of galaxies.  The Next
Generation Very Large Array (ngVLA) has the potential to measure
proper motions for large numbers of active galactic nuclei in nearby
clusters and groups of galaxies.  Given sufficient angular resolution,
radio data can also provide precise, direct geometric distances to a
small number of galaxies with either expanding supernova shells or
with bright masers in disks.  Excitingly, around the time we
anticipate the ngVLA to be commissioned, we also anticipate that LISA
will launch, opening up the era of standard siren distances to nearby
galaxies which can be used to obtain precise distances to many
galaxies, and also to provide precise calibration of the distances to
the others.

Understanding the spatial distribution of dark matter in clusters of
galaxies is one of the keys to understanding structure formation.
Probes of dark matter halos typically rely on observables that are
integrated along the line of sight (e.g. gravitational lensing shear,
X-ray and Sunyaev-Zel'dovich surface brightness profiles), which must
be transformed modulo physical assumptions to yield a mass
distribution.

Collecting ``test masses'' within clusters of galaxies with
well-specified positions and three-dimensional velocities can give
precision information about the gravitational potential at specific
locations within the potential.  This information can then be combined
with information collected through other techniques to eliminate
degeneracies associated with line-of-sight projection and provide a
far more precise determination of the mass distribution than is
currently possible.  The above technique is precisely the approach
taken in measuring the mass distribution of the Milky Way. The
observations described below would probe the mass distribution of
Virgo at nearly the fidelity that has been possible for the Milky Way
prior to Gaia, taking a first key step towards a detailed mass model
for the Virgo cluster.

Moreover, measuring proper motions of galaxies is of great value for
understanding a variety of other issues related to understanding
galaxy evolution and cluster evolution.  Precision proper motions of
galaxies can be measured via masers in ordered disks, for the galaxies
where they are found (Reid et al. in this volume; see also
\citet{Reid09}).  Here we focus on alternative tracers of the
galaxies' positions that can be used at large distances, even for
galaxies without masers.

\section{Anticipated results}

Galaxies within clusters already have their positions on the sky and
their radial velocities well measured.  This information is sufficient
to obtain estimates of the mass enclosed within a given radius based
on some assumptions about how the clusters de-project themselves.  To
improve upon these results, precise distances and proper motions are
necessary.  We can expect in the next 20 years to have a good set of
standard siren distances for many galaxies in the Virgo Cluster, and
perhaps a few galaxies in the Fornax Cluster.  Obtaining the proper
motions requires the combination of sensitivity and angular resolution
of the ngVLA, and benefits strongly from having significant
sensitivity on baselines well in excess of 300 km.

The Laser Interferometric Gravitational-wave Antenna will detect
double black hole binaries within its period range as continuously
emitting sources of gravitational waves.  From the gravitational waves
themselves, it should be possible to measure standard siren distances
to a few percent, and to identify the host galaxy from within an error
volume of about $1~{\rm Mpc}^3$ or less. Once the host galaxy has been
correctly identified, its location can used to refine the distance
estimate, bringing the error in distance down to about 1\%.  This will
rely on having small gravitational accelerations of the merging black
holes; however, this is expected, because even in models where they
form in globular clusters, they are expected to be ejected
\citep{Kremer18}.  Furthermore, for the shortest period binaries in
the LISA band, the cluster acceleration will affect the measurement of
the period derivative by only a few percent.

Obtaining proper motions of the host galaxies of the gravitational
wave sources then fills out the six-dimensional phase space
measurements.  The velocity dispersion of galaxies in the Virgo
Cluster is about 700 km/sec, while the velocity dispersions within the
more massive galaxies can be $\sim300-400$ km/sec.  As a result, the
use of optically bright tracers (e.g. red giants, planetary nebulae or
globular clusters) to measure the proper motions of the massive
galaxies that are most likely to contain the binary black holes
providing standard siren distances is problematic -- even with precise
measurements of the velocities of individual objects, several thousand
of them would be required to centroid the broad velocity dispersions
to 10 km/sec precision.  While some of the largest galaxies in the
Virgo Cluster have thousands of globular clusters, some fraction of
these (especially in M87) are likely to be intracluster, intergalactic
globular clusters projected against the host galaxy, and many of the
clusters, whether members or not, will be too faint for precise proper
motion measurements, even with JWST.  Furthermore, the uncertainties
on these measurements will not improve substantially with a long time
baseline, because the primary source of uncertainty will be how well
the velocity distribution has been sampled, rather than the
uncertainties on the individual measurements.  Given that the standard
siren distances will be accurate to a few percent, it is of great
value to obtain similar precision in the proper motions in order not
to have them dominate the total uncertainty budget for the phase space
measurements.  Making these measurements thus requires the use of the
galaxies' nuclei, rather than their stellar populations.

A few of the nearby groups of galaxies are also excellent candidates
for studies of dark matter halos on different scales.  \citep{Oehm17}
have shown that proper motion dispersions of the centers of mass of
the galaxies of tens of microarcseconds per year, corresponding to
hundreds of km/sec are expected among the galaxies in the M81 group,
and that the relative magnitudes of these space velocities can provide
tests of whether dynamical friction is happening as predicted by dark
matter theory \citep{Kroupa15}.  Here, the smaller galaxies, M82 and
NGC~3077 may not contain bright enough active galactic nuclei for
estimating their proper motions, but as dwarf galaxies, with small
internal velocity dispersions, other radio sources may be effective
tracers of their motions.  These tracers could include planetary
nebulae and the brightest HII regions, and, in some cases, individual
luminous blue variable stars, through their wind emission.

Finally, at a lower precision level, merging clusters and merging
groups of galaxies can be studied to determine the relative merger
speeds, and hence to determine the ratio between the masses of the two
clusters.  The Bullet Cluster \citep{Clowe06}, for example, has
easily detectable AGN in both subclusters \citep{Malu16}, and with
a relative velocity of 4500 km/sec at $z=0.296$, would be expected to
show a space velocity of about 1 $\mu$arcsec/year, as would Abell
2034, which has a merger speed which is somewhat slower, but which is
located more nearby.  The Bullet Cluster itself is too far south for
ngVLA.  In the near future, with $eROSITA$, $WFIRST$ and $Euclid$, we
can expect new merging clusters and merger compact groups of galaxies
to be discovered, allowing for precise tests of the extremes of the
dynamic intermediate redshift Universe.

\section{Ancillary astrophysics: understanding the morphology-density correlation in clusters}

In addition to enabling modelling of the dark matter distribution,
proper motions for cluster galaxies provide an incredibly powerful
tool for disentangling the physical factors impacting the evolution of
these galaxies. In particular, it has long been known that spiral
galaxies are quite rare in dense clusters of galaxies, especially in
their cores \citep{Oemler74}.  The subset of spirals that do exist
in cluster cores therefore provide an important sample for
understanding in detail the processes by which these galaxies are
quenched. These galaxies are most likely on their first approach into
the cluster, but full 6-D phase space information is necessary to
determine their orbits, locations relative to the cluster center, and
full orbital velocities. All of these quantities are critical for
understanding the impact of cluster-specific quenching processes. For
instance, quantifying the impact of present (future) ram pressure
requires knowledge of the orbital velocity (and orbit).  Such
information currently does not exist -- proper motions are required.

Even for galaxies that lack standard siren distance measurements, full
3-D velocities and angular positions combined with a model for the
mass distribution are sufficient to identify substructures and
constrain orbital motions at a fidelity that enables one to
disentangle whether a galaxy is infalling for the first time, is
associated with an accreted substructure, or had resided in the
cluster core for Gyrs. Consequently, it becomes possible to infer the
distribution of environments that galaxy has encountered within the
cluster.

\section{Limitations of current astronomical instrumentation}

The instrument best suited to address this problem at the current time
would be the VLBA.  It has 0.185 mJy noise in one hour at 40 GHz and
0.2 masec resolution.  This leads to two key disadvantages relative to
the ngVLA.  First, the exposure times start to become prohibitive
(although some long observations to provide a first epoch at the
present time, so that the ngVLA has a long time baseline for proper
motions would be valuable).  Second, the systematics with the VLBA are
likely to be significantly worse than with ngVLA if the Long Baseline
Major Option is implemented; systematics increase with the separation
between the source and the phase calibrator, and with approximately
ten times better sensivity the angular separation of the nearest good
calibrator will typically be much smaller.  Optical telescopes can
obtain the combination of signal-to-noise and resolution nominally
needed to address these questions, but then rely strongly on extremely
precise flat-fielding and modelling of the point spread function that
goes well beyond what has currently been established to be possible.
With Contintental baselines, relative astrometry of AGN to
microarcsecond accuracy per epoch in the radio should be
straightforward to achieve, while optical astrometry appears to be
limited by systematics at a level roughly 10-100 times worse.

\section{Connection to unique ngVLA capabilities}

As discussed above, this work can be done only with radio-based proper
motion measurements of the active galactic nuclei of Virgo Cluster
galaxies.  The excellent sensitivity of the ngVLA is needed to make
these measurements at high frequency where the systematics due to core
shifts are least important, and with the ability to use moderately
bright phase calibrators located very close to the source.  Some
aspects of this work can be done with the Southwest configuration for
the ngVLA, but many of these projects would require, or at least
strongly benefit from,

\section{Experimental layout}
\subsection{Virgo Cluster}
The radio flux densities of the large galaxies in the Virgo Cluster
are mostly $\sim 1$ mJy or more \citep{Merloni03}.  We take
this as a baseline value.  For an AGN at this brightness, 
with the ngVLA we can expect a noise level of about 0.3 $\mu$Jy in one
hour with the ngVLA, working at 40 GHz with a beam size of 2
milliarcseconds.  For a 1 mJy source, this then gives signal to noise
of about 3000, meaning that the positional accuracy of the AGN will be
good to about 0.7 $\mu$arcsec, which is likely to be approximately the
systematic limit.  Obtaining a better handle on the systematics, so
that the statistical limits can be reached, provides a strong
motivation here for the Long Baseline Major Option; still, taking more
observations and over a longer time baseline can also be done to help
ensure that the results are robust.  Positions measured to this
accuracy will yield proper motions accurate at the level of $\approx$
50 km/s on a one-year time baseline, so that within 5 years, proper
motions could be measured to an accuracy of 10 km/sec.  Because the
observations needed are quite short this also allows for extra checks
to be done to ensure that systematics such as core shifts are not
problematic, by allowing extra observations at different frequencies,
and observations over a range of source fluxes.  Additionally, use of
longer baselines would improve both the systematics and the
sensitivity.

While only the galaxies with LISA binaries are likely to be useful for
the six-dimensional phase space measurements, even having good
five-dimensional phase space measurements for all the large galaxies
will provide very valuable information.  We thus believe it would make
most sense to observe the 50 or so most luminous galaxies.  This also
has the advantage of not requiring that the LISA detections have
already been made at the time the ngVLA starts collecting data.  For
astrometric projects, given the importance of having a long time
baseline, this is vital.  We thus expect that conducting this project
would require about 50 hours per epoch to observe the 50
radio-brightest galaxies.  It may occur that LISA uncovers some
stellar-mass binary black holes in fainter galaxies, which may lead to
a bit of extra time being required, but as a first order estimate, we
can expect this program to require about 250 hours of ngVLA time,
spread over five years.

\subsection{M81 group}

For the M81 group, the test of whether dark matter is producing
dynamical friction proposed by \citet{Oehm17} relies on making
precise proper motions of M81, M82 and NGC 3077, and precision of 2
$\mu$sec/year is needed.  To obtain such precision over a 5 year
baseline, one needs precision of 7 $\mu$sec per measurement (slightly
worse precision can be tolerated using more than 2 measurements).

\subsection{Merging clusters}

As a template object, we consider Abell~2034.  This is a merging
cluster of glaaxies with an estimated merger speed of about 2000
km/sec \citep{Owers14}, mostly across the plane of the sky
\citep{Monteiro18}.  The cluster is at a redshift of $z=0.115$, or a
distance of 433 Mpc.  The proper motion should thus be $0.8
\mu$arcsecond per year.  Thus, with systematics that should be
achievable in a straightforward manner with the Long Baseline Major
Option, it should be possible to measure the proper motion at
$10\sigma$ in a little over a decade.  In fact, it is not yet certain
if this cluster is in the early or late phases of its merger at the
present time \citep{Monteiro18}.  At the present time, only relatively
low angular resolution radio images have been obtained for this
galaxy.  It is likely, though that there should be at least a few AGN
with flux densities above 50 $\mu$Jy in each sub-cluster, allowing for
micro-arcsecond positional accuracy in reasonable exposure times.
This system is clearly the best case right now for a merger cluster
which is near enough, with fast enough tangential velocities, and far
enough North for proper motions to be useful.  It is likely, though,
that in the near future, new surveys will uncover some additional
candidates.

\acknowledgements We are extraordinarily grateful to Matt Benacquista for valuable discussions and providing some of the foundational ideas that led to this work and to Joan Wrobel, Pau Amaro-Seaone and Robyn Sanderson for additional useful discussions.



\end{document}